# Doping Induced Second Harmonic Generation in Centrosymmetric Graphene from Quadrupole Response

Yu Zhang,[1] Di Huang,[1] Yuwei Shan,[1] Tao Jiang,[1] Zhihong Zhang,[2] Kaihui Liu,[2] Lei Shi,[1,3] Jinluo Cheng,[4] John E. Sipe,[5] Wei-Tao Liu,[1,3*] Shiwei Wu[1,3*]

[1] State Key Laboratory of Surface Physics, Key Laboratory of Micro and Nano Photonic Structures (MOE), Department of Physics, and Institute for Nanoelectronic Devices and Quantum Computing, Fudan University, Shanghai 200433, China

[2] State Key Laboratory for Mesoscopic Physics and School of Physics, Peking University, Beijing 100871, China

[3] Collaborative Innovation Center of Advanced Microstructures, Nanjing 210093, China

[4] Changchun Institute of Optics, Fine Mechanics and Physics, Chinese Academy of Sciences, Changchun, Jilin 130033, China

[5] Department of Physics, University of Toronto, Toronto, Ontario, Canada

**Abstract:**

For centrosymmetric materials such as monolayer graphene, no optical second harmonic generation (SHG) is generally expected because it is forbidden under the electric-dipole approximation. Yet we observed a strong, doping induced SHG from graphene, with its highest strength comparable to the electric-dipole allowed SHG in non-centrosymmetric 2D materials. This novel SHG has the nature of an electric-quadrupole response, arising from the effective breaking of inversion symmetry by optical dressing with an in-plane photon wave vector. More remarkably, the SHG is widely tuned by carrier doping or chemical potential, being sharply enhanced at Fermi edge resonances, but vanishing at the charge neutral point that manifests the electron-hole symmetry of massless Dirac Fermions. The striking behavior in graphene, which should also arise in graphene-like Dirac materials, expands the scope of nonlinear optics, and holds the promise of novel optoelectronic and photonic applications.

Second harmonic generation (SHG) is the most fundamental second-order nonlinear optical process, described by $\mathbf{P}(2\omega = \omega + \omega) = \chi^{(2)}(\omega, \mathbf{q}):\mathbf{E}(\omega)\mathbf{E}(\omega)$ [1]. In this process, the output signal is frequency doubled from the incident photon field of $\mathbf{E}(\omega)$. Here, $\chi^{(2)}(\omega, \mathbf{q})$ is the rank-three nonlinear susceptibility tensor and depends on the incident frequency $\omega$ and photon wave vector $\mathbf{q}$. Since $\mathbf{q}$ is typically small, a Taylor expansion yields $\chi^{(2)}(\omega, \mathbf{q}) \approx \chi^{(2)}_{ED}(\omega) + \chi^{(3)}_{EQ}(\omega)\mathbf{q} + o(\mathbf{q^2})$. $\chi^{(2)}_{ED}(\omega)$ is the leading electric-dipole (ED) term, and $\chi^{(3)}_{EQ}(\omega)$ is the often neglected electric-quadrupole/ magnetic-dipole term [1,2] or the EQ response for simplicity. For the electric-dipole allowed SHG to exist, the breaking of inversion symmetry is essential. Hence, SHG is a sensitive probe to symmetry-governed phenomena such as ferroelectricity [3], valley pseudospin [4], and phase transitions [5].

For 2D materials such as hexagonal boron nitride, transition metal dichalcogenide and monochalcogenide, their atomic lattices in monolayer form are non-centrosymmetric, giving rise to the electric-dipole allowed SHG. In fact, SHG has become an indispensable tool to characterize their crystal orientation, stacking symmetry and electronic features [6-10]. Yet for an isolated monolayer graphene, it is centrosymmetric and no electric-dipole SHG is allowed. The third-order optical nonlinearity such as third-harmonic generation [11-15] and four-wave mixing [14,16] was often regarded as the dominant nonlinear process in graphene. Only weak SHG was observed on supported monolayers, which was attributed to the inversion symmetry breaking by the underlying substrate [17] or an in-plane electric current [18,19]. Therefore, graphene provides a unique platform to study unusual SHG responses such as valley polarization induced SHG [20,21] and the EQ response beyond the electric-dipole approximation [22,23], as theoretically proposed.

In this work, we exclusively investigate the EQ response of SHG in graphene by introducing an in-plane photon wave vector **q** at oblique incidence, which effectively break the overall inversion symmetry of the system [22-25]. By comparing with the SHG response at normal incidence, we could exclude other possible origins such as the substrate-induced SHG [17]. More importantly, we find that the EQ response of SHG in graphene is widely tunable by carrier doping or chemical potential and exhibits strong Fermi edge resonances, by using ion-gel electric gating [14]. The strong EQ-SHG in graphene is even comparable to the ED allowed SHG from non-centrosymmetric 2D materials with parabolic bands [6-8,10], and the effective nonlinear susceptibility is about four orders of magnitude stronger than the EQ response in bulk fused silica [26] (rarely reported for other materials). Furthermore, we find that this EQ process is intrinsically sensitive to the electron-hole symmetry of the energy bands, strictly vanishing at zero chemical potential. In the past, nonlinear optics have long been established as powerful tools for analyzing the space and time symmetries of materials [1,5-10]. Now we can add the electron-hole symmetry to this family, and utilize the EQ response to study related Dirac materials such as topological insulators [27], Dirac and Weyl semimetals [28].

Our experimental geometry is sketched in Fig. 1(a), with the excitation beam ($\hbar\omega$ = 0.95 eV, unless otherwise noted) incident at 45$^o$ from the surface normal of the ion-gel gated graphene [14]. The doping level or chemical potential in graphene could be electrically tuned by the gate voltage and the relationship between gate voltage and chemical potential was determined from the infrared transmittance spectra. Additional experimental details are provided in Ref. 14 as well as in Supplemental Material [29]. Along the direction of specular reflection, a signal at $2\omega$ could be observed from graphene, which is vanishing with the chemical potential tuned toward the charge neutral point (CNP), but grows rapidly upon the increase of chemical potential $|\mu|$ [Fig. 1(b) and 1(c) for S-polarized incidence, and Fig. 1(d) and 1(e) for P-polarized incidence]. In comparison, there is no observable SHG on ion-gel covered fused silica substrate at any gate voltage [black lines in Fig. 1(c) and Fig. 1(e)]. The signal intensity on graphene was proportional to the square of incident fluence [red symbols in Fig. 1(f)], confirming it to be an SHG response.

In earlier studies by Dean *et al.* [17], weak SHG was observed from pristine graphene monolayers at oblique incidence of $60^o$, and was attributed to the usual ED-type response due to the breaking of inversion symmetry by the presence of the oxidized silicon substrate. Since ED-SHG and EQ-SHG arise from different susceptibility tensors, they can be distinguished by varying the experimental geometry from the oblique incidence to normal incidence. If the SHG were purely of ED-type, the signal at normal incidence should be comparable to that under the oblique geometry. If the EQ-SHG dominates, a large intensity contrast between the two geometries would be expected, because an in-plane component of photon wave vector is essential to the EQ-SHG (see details in SM [29]). We thus performed the SHG measurement under the normal incidence [Fig. 2(a)]. The signal is only about 0.6% of the oblique SHG at the same chemical potential [Fig. 2(b)], consistent with the dominance of EQ-SHG.

The EQ-SHG originates from the nonlinear polarization $\mathbf{P}_{EQ}(2\omega) \propto \chi_{EQ}^{(3)} \mathbf{q} \mathbf{E}(\omega)^2$. This relation imposes symmetry restrictions to the EQ-SHG, which can be manifest in polarization features as displayed in Fig. 3. Because of the 2D geometry of graphene, we consider only the in-plane subset of $\chi_{EQ}^{(3)}$ (full expression given in SM [29]). Since graphene belongs to the point group $D_{6h}$ (*6/mmm*), this subset has an identical construction to that of an isotropic surface [1]; in other words, the graphene plane is optically isotropic for $\chi^{(3)}$ processes [14,31]. When the incident beam is P-polarized, the incident plane defines a mirror plane of the system [Fig. 3(a)]. So $\mathbf{P}_{EQ}(2\omega)$, which is a polar vector, must lie in the incident plane; otherwise it would change sign after the mirror reflection and break the symmetry. When the incident beam is S-polarized, the $\mathbf{E}(\omega)^2$-relation in the expression of $\mathbf{P}_{EQ}(2\omega)$ again imposes a mirror symmetry along the incident plane, restricting $\mathbf{P}_{EQ}(2\omega)$ to remain in this plane [Fig. 3(b)]. Hence the EQ-SHG must be P-polarized in both cases [22]. We verified that by changing the angle ($\theta$) between the signal analyzer and the beam incident plane. As seen in Fig. 3(c) and 3(d) (red squares), for both P- and S-polarized excitations, the signal polarization patterns can be nicely fitted by the function $\cos^2\theta$, corresponding to a linearly P-polarized output.

On the other hand, if the incident beam has both P and S components, the EQ-SHG is not a trivial superposition of two P-polarized fields, but in general becomes elliptically polarized [22,23,25] [Fig. 3(e), red squares]. This is because there is no longer any mirror plane defined by $\mathbf{q}$ or $\mathbf{E}(\omega)$, so $\mathbf{P}_{EQ}(2\omega)$ is not restricted to be along either of them. These behaviors are in contrast to the ED-type third harmonic generation (THG) arising from $\mathbf{P}_{ED}(3\omega) \propto \chi_{ED}^{(3)}\mathbf{E}(\omega)^3$, which always has a mirror plane defined along $\mathbf{E}(\omega)$, and keeps $\mathbf{P}_{ED}(3\omega)$ parallel to the in-plane projection of $\mathbf{E}(\omega)$ [Fig. 3(c)-3(e), gray dots] [14,31]. Detailed calculations based on EQ and ED types of $\chi^{(3)}$ tensors yielded nice fits to the polarization patterns of SHG and THG, respectively [solid lines in Fig. 3(c)-3(e)] (see SM for details [29]).

We now discuss the prominent $\mu$-dependence of the EQ-SHG [Fig. 4(a)]. The very weak SHG near $\mu = 0$ or CNP grows rapidly as $\mu$ approaches the Fermi edge resonances at one-photon ($2|\mu| = \hbar\omega$) and two-photon ($|\mu| = \hbar\omega$) energies [Fig. 4(b)]. Meanwhile, the S- and P-polarized excitations result in distinct line profiles [Fig. 4(a)], indicating that different tensor elements have different $\mu$-dependences. Throughout, the intensity could be tuned over two orders of magnitudes by electrically tuning the chemical potential $\mu$. Theoretically, the $\mu$-dependence of EQ-SHG has been predicted by the full quantum mechanical calculations by Wang *et al.* [22] and Cheng *et al.* [23], and is attributed to the resonant transitions in linearly dispersed band structure of graphene. As shown in Fig. 4(c), the calculation nicely reproduces the Fermi-edge resonances, as well as the relative strength between different tensor elements observed in our measurements [Fig. 4(a)]. The calculated magnitude of the effective second-order nonlinear susceptibility $|\chi_{EQ}^{(3)}q_x|$ is also close to the experimental values.

As previously reported [14], the third harmonic generation also grows with chemical potential $\mu$ and exhibits the Fermi edge resonances (Fig. S1). On the other hand, there is a marked difference between the $\mu$-dependence of EQ-SHG and ED-THG. In contrast to the vanishing EQ-SHG at $\mu = 0$, the THG is readily detectable as shown in Fig. 1(c) and 1(e). Accordingly, under the two-band approximation, the calculation shows that $\chi_{ED}^{(3)}$ for THG and four-wave mixing are even functions of $\mu$ [31], but $\chi_{EQ}^{(3)}$ for EQ-SHG is an odd function with respect to $\mu$ that has to vanish at CNP [22,23] [Fig. 4(d)-4(f)]. The latter is a direct consequence of the spatial dispersion for EQ-type responses, in combination with the electron-hole symmetry and time-reversal symmetry of graphene. The detailed proof is provided in the SM [29], and here we briefly sketch the physical picture. Each element of the optical susceptibility tensor is a summation of diagrams like the one in Fig. 4(g), which depicts a transition with a photon wave vector $\mathbf{q}$ on the right-hand-side of the energy bands at a finite $\mu$. Due to the electron-hole symmetry and time reversal symmetry, the diagram is equivalent to a transition with $-\mathbf{q}$ on the left-hand-side ($\mathbf{k} \to -\mathbf{k}$) for $\mu \to -\mu$. Therefore, the susceptibility tensor element at $\mathbf{q}$ and $\mu$ is equivalent to that at $-\mathbf{q}$ and $-\mu$. For the EQ-SHG, each diagram contains a matrix element $\langle \mathbf{q} \rangle$; and since $\langle -\mathbf{q} \rangle = -\langle \mathbf{q} \rangle$, we then have $\chi_{EQ}^{(3)}(\mu, \mathbf{q}) = \chi_{EQ}^{(3)}(-\mu, -\mathbf{q}) = -\chi_{EQ}^{(3)}(-\mu, \mathbf{q})$, which has an odd parity with respect to $\mu$. This is fundamentally different from susceptibility for ED-THG, which only involves matrix elements of dipole moments, thus will not change sign upon the inversion of electron and hole bands.

Finally, we comment on the strength of EQ-SHG in graphene. In conventional materials, the EQ-SHG is often eclipsed by the ED-type counterpart. For example, the effective nonlinearity of the EQ response in fused silica is only about $4\times10^{-3}$ pm/V [26]. Yet it can be exceptionally strong in graphene, reaching $|\chi^{(3)}_{EQ,xxyy}q_x|$~30 pm/V upon Fermi edge resonances for $\hbar\omega = 0.95$ eV. This is comparable to the ED-allowed $|\chi^{(2)}_{ED}|$ in non-centrosymmetric 2D materials such as hBN monolayer [8] and ABA-stacked trilayer graphene [9], which is typically about 10-100 pm/V. For TMDC monolayers such as MoS$_2$, the above $|\chi^{(2)}_{ED}|$ can increase by an order upon excitonic resonance [6,8,10] (Fig. S2). Nonetheless, such resonance is always accompanied by enhanced absorption and lowered damage threshold. In contrast, the Fermi-edge resonances in graphene actually reduce the absorption, allowing graphene to be pumped at an even higher fluence for stronger responses. Moreover, as pointed out in Ref. 22, the dipole transition matrix element scales with $\omega^{-1}$ for the massless Dirac fermions in graphene, instead of $\omega^{-1/2}$ for conventional materials with the parabolic energy dispersion. We thus expect the EQ-type responses of graphene to have an even higher efficiency in the infrared frequency range. All these may explain the strong surface plasmon enhanced difference-frequency generation observed by Constant *et al.* [32] and Yao *et al.* [33]. Thus the combination of electrically tunable strong response and suppressed absorption damage makes graphene a unique and highly promising candidate in nano-photonic and optoelectronic applications, ranging from optical rectification for terahertz generation and sum-frequency generation for parametric conversion.

In conclusion, we reveal in this work a doping-induced, strong SHG response from the centrosymmetric graphene monolayer, which has a comparable strength to that from non-centrosymmetric 2D materials. Based on its $\mu$-dependence, symmetry properties, and incident angle dependence, we attribute it to the EQ-type of response resultant from the unique properties of massless Dirac fermions. Interestingly, we find this EQ response is intrinsically sensitive to the electron-hole symmetry of the band structure, becoming strictly zero at the charge neutral point. The understanding derived here is readily applicable to other related Dirac materials such as topological insulators [27], Dirac and Weyl semimetals [28]. Therefore, graphene provides a unique platform for investigating unusual nonlinear optical phenomena, which not only can expand the horizon of nonlinear optics in quantum materials, but also has a large potential in novel device applications [33,34].

**Acknowledgements:** We thank Prof. Yuen-Ron Shen for insightful discussion. The work at Fudan University was supported by the National Basic Research Program of China (Grant No. 2014CB921601), National Key Research and Development Program of China (Grant Nos. 2016YFA0301002, 2016YFA0300900), National Natural Science Foundation of China (Grant Nos. 91421108, 11622429, 11374065), and the Science and Technology Commission of Shanghai Municipality (Grant No. 16JC1400401). Part of the sample fabrication was performed at Fudan Nano-fabrication Laboratory. K.L. is supported by National Natural Science Foundation of China (Grant No. 51522201). J.C. is supported by National Natural Science Foundation of China (Grant No. 11774340). J.E.S. is supported by the Natural Sciences and Engineering Research Council of Canada. Y.Z. and D.H. equally contributed to this work.

* Corresponding authors: swwu@fudan.edu.cn, wtliu@fudan.edu.cn

**Figures and Captions**

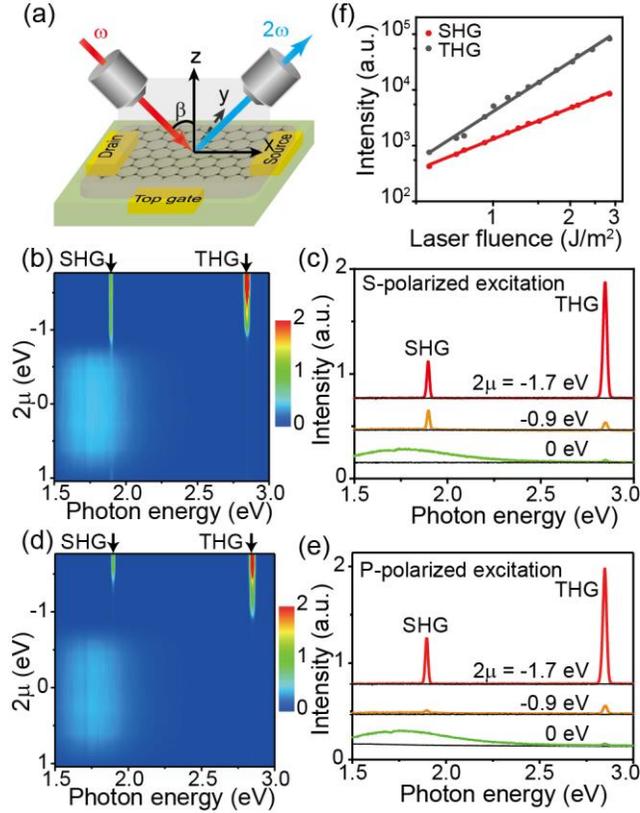

FIG. 1. SHG from ion-gel gated graphene with oblique incident excitation. (a) Experimental geometry at oblique incidence of 45°. The graphene supported on a fused silica substrate is covered with ion-gel for gating, through which the chemical potential $\mu$ could be tuned from -0.9 eV to 0.5 eV. The laser excitation and signal collection are done through microscope objectives with numerical apertures of 0.35 and 0.40, respectively. The incident photon energy was 0.95 eV, unless otherwise noted. (b) Specularly reflected spectra with S-polarized excitation as a function of $2\mu$. (c) Selected spectra in (b) at different chemical potential $\mu$. The spectra are offset for clarity. (d), (e) Corresponding spectra with the P-polarized excitation. The spectra from bare fused silica covered with ion-gel at corresponding gate voltages are also shown as black lines in (c) and (e) for comparison. The broadband background at low doping is from the ultrafast nonlinear photoluminescence, which could be switched off at high doping [35-36]. The incident laser fluence was 2.4 J/m$^2$. (f) Incident fluence dependence of SHG (red) and THG (gray)

intensities at heavily doped level ($\mu \approx -0.88$ eV) on a double-logarithmic scale. Solid lines are fits with slopes of $1.85 \pm 0.02$ and $2.95 \pm 0.06$, respectively.

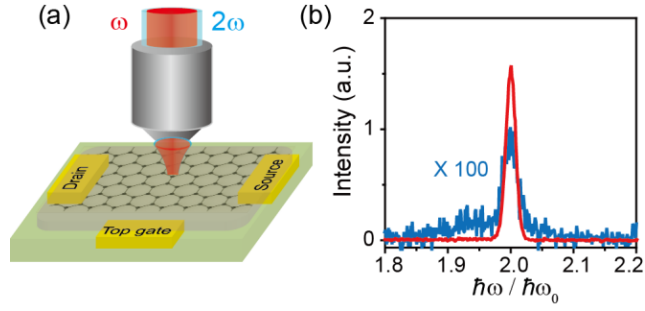

FIG. 2. Comparison of SHG between normal and oblique incidences. (a) Experimental geometry at normal incidence. The excitation and collection are through a same objective with numerical aperture as high as 0.95. (b) SHG spectra from heavily doped graphene ($\mu$ = -0.8 eV) at the same incident fluence of 2.4 J/m$^2$. The red and blue data represent the cases of oblique (S-polarized) and normal incidences, respectively. The spectrum at normal incidence is magnified by 100 times.

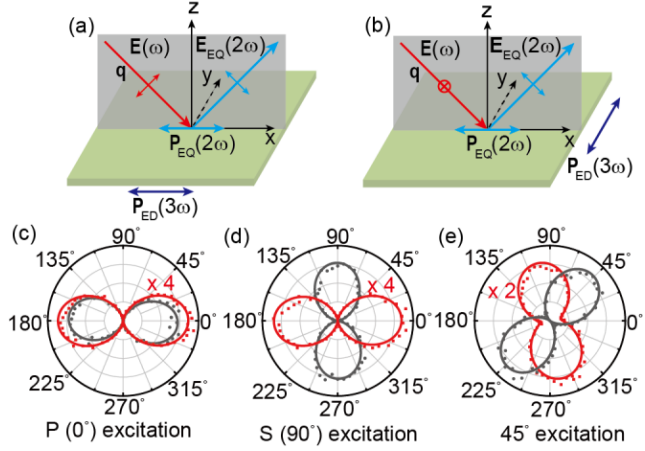

FIG. 3. Polarization characteristics of the SHG at oblique incidence of 45°. (a), (b) Symmetry properties of the overall system defined by both the input electric field and wave vector. For both S- and P-polarized excitations, $\mathbf{P}_{EQ}(2\omega)$ is confined in the mirror plane defined by the incident plane, while $\mathbf{P}_{ED}(3\omega)$ follows the in-plane projection of $\mathbf{E}(\omega)$. (c)-(e) The SHG (red dots) intensity versus the angle $\theta$ between the beam incident plane and analyzer polarization direction, under at S-, P-, and 45°-polarized excitations, respectively. The THG (gray dots) data are also shown for comparison. Solid lines are fits using equations in SM [29]. The graphene was heavily doped at μ = -0.88 eV.

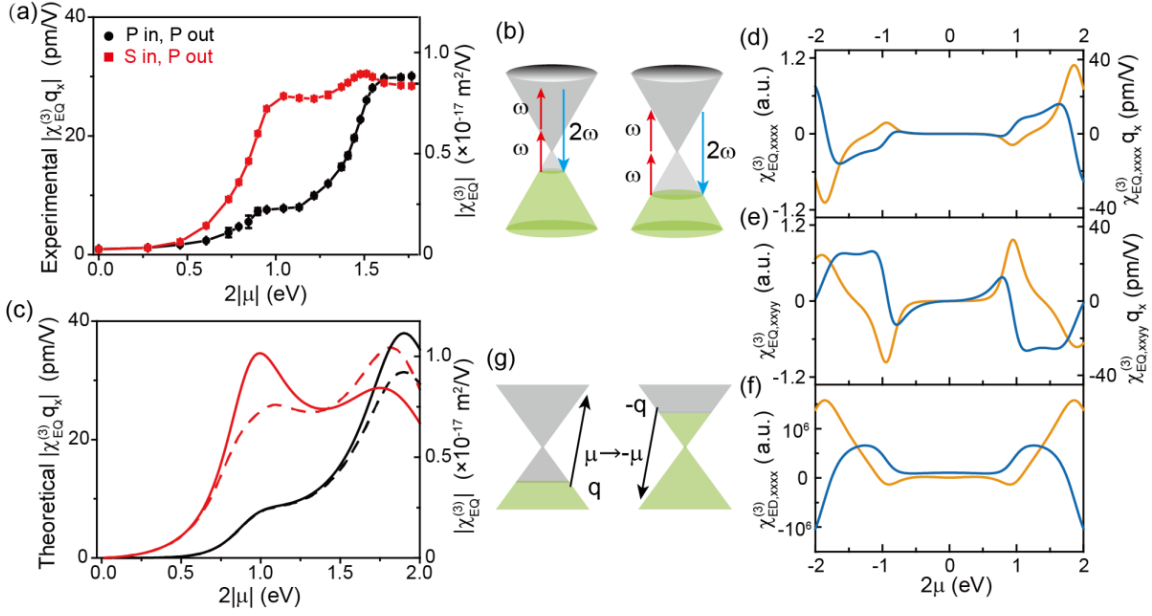

FIG. 4. Chemical potential dependence of the oblique SHG. (a) The extracted second-order nonlinear susceptibility $|\chi^{(3)}_{EQ} q_x|$ of the SHG under S- (red dots) and P- (black dots) polarized light excitation versus chemical potential $2|\mu|$. The data were extracted from Fig. 1(b) and Fig. 1(d) for negative chemical potentials. (b) Transition diagrams for one- and two-photon Fermi-edge resonances, respectively. (c) Theoretically calculated $|\chi^{(3)}_{EQ,xxyy} q_x|$ (red) and $|\chi^{(3)}_{EQ,xxxx} q_x|$ (black) for S- and P-polarized excitations, respectively. The dashed and solid lines are from the theories by Wang *et al.* [22] and Cheng *et al.* [23], respectively. In the calculation, the temperature T = 300 K (ignoring the effect of high electron temperature for $2|\mu| \lesssim \hbar\omega$) and the resonant damping factor $\Gamma = 0.2 \times |\mu|$ eV are used [14]. (d), (e) Real (yellow) and Imaginary (blue) part of $\chi^{(3)}_{EQ,xxxx}$ and $\chi^{(3)}_{EQ,xxyy}$, corresponding to EQ-SHG under P- and S-polarized excitations, respectively. On the right vertical axis, the effective second-order nonlinear susceptibilities $\chi^{(3)}_{EQ,xxxx} q_x$ and $\chi^{(3)}_{EQ,xxyy} q_x$ are also shown. (f) Real (yellow) and Imaginary (blue) part of $\chi^{(3)}_{ED,xxxx}$, corresponding to ED-THG. The calculations were followed from the theory by Cheng *et al.* [23, 31]. (g) Transition schematics under the operation of $\mu \to -\mu$ and $\mathbf{k} \to -\mathbf{k}$ with the electron-hole symmetry and time reversal

symmetry for $\mathbf{q} \to -\mathbf{q}$, where $\mathbf{k}$ is the electron wave vector and $\mathbf{q}$ the photon wave vector.

Supplemental Material for

# Doping Induced Second Harmonic Generation in Centrosymmetric Graphene from Quadrupole Response

Yu Zhang, Di Huang, Yuwei Shan, Tao Jiang, Zhihong Zhang, Kaihui Liu, Lei Shi,

Jinluo Cheng, John E. Sipe, Wei-Tao Liu, Shiwei Wu

**The Supplemental Material includes:**

**Supplemental Texts 1-5.**

**Supplemental Table S1.**

**Supplemental Figures S1-S3.**

**Supplemental Texts**

**1. Experimental details**

**a) Device fabrication**

Single crystalline monolayer graphene [30] used in the experiment was grown by chemical vapour deposition (CVD) and transferred onto optically transparent fused silica substrates. Source, drain and gate electrodes (50-nm Au and 5-nm Cr) were patterned through a dry stencil mask by electron beam deposition. All the electrodes were wire-bonded to a chip carrier for electrical control. Ion-gel gating was achieved by uniformly applying freshly prepared ion-gel solution ([EMIM][TFSI] and PS-PEO-PS in dry dichloromethane) onto the graphene devices, and further drying in a glove box filled with high purity argon gas. The charge neutral point of graphene was determined by its maximum resistance in response to the gate voltage. The relationship between the applied gated voltage and the chemical potential in graphene was experimentally determined by measuring the infrared transmittance spectra of gated graphene, as described in Ref. 14.

**b) Optical measurement**

The optical microscopic setup to measure the SHG from gated graphene at oblique incidence is shown in Fig. S3. In this setup, a femtosecond optical parametric oscillator (OPO, ~200 fs pulse duration, 80 MHz repetition rate) with tunable wavelength was used as the excitation source. The linearly polarized light was focused onto the sample at 45° incident angle by a 20× long working-distance objective (NA = 0.35). The reflected SHG signal was collected by a 50× long working-distance objective (NA = 0.40), and guided to a fiber-coupled spectrograph equipped with a liquid nitrogen cooled silicon charge-coupled device to acquire spectrum or a single-photon counting silicon avalanche photodiode to acquire intensity. During the whole measurement, the graphene device was maintained in a dry nitrogen environment at room temperature. The experimental results were confirmed and reproduced on multiple sample positions and different graphene devices.

## 2. The ED-SHG and EQ-SHG under different excitation geometries

### a) The tensor elements for EQ-SHG and ED-SHG

An isolated graphene monolayer belongs to centrosymmetric point group $D_{6h}$ ($6/mmm$); a supported graphene monolayer on an isotropic surface may experience asymmetry induced by the substrate, which reduces the symmetry to the non-centrosymmetric point group $C_{6v}$ ($6mm$). In both cases there is a $C_6$ axis perpendicular to the lattice plane, and three mirror planes along the arm-chair direction. In the crystalline coordinates ($a$, $b$, $c$), we set $c$ parallel to the $C_6$ axis, and $b$ along one of the arm-chair directions with $a$ perpendicular to $b$. $\overleftrightarrow{\chi}^{(3)}_{EQ}$, representing the EQ-SHG process, for the two point groups is identical, which has the following nonzero in-plane tensor elements [1]:

$$\chi^{(3)}_{aaaa} = \chi^{(3)}_{bbbb} = \chi^{(3)}_{aabb} + \chi^{(3)}_{abab} + \chi^{(3)}_{abba}, \begin{cases} \chi^{(3)}_{aabb} = \chi^{(3)}_{bbaa} \\ \chi^{(3)}_{abab} = \chi^{(3)}_{baba} \\ \chi^{(3)}_{abba} = \chi^{(3)}_{baab} \end{cases}, \begin{cases} \chi^{(3)}_{abab} = \chi^{(3)}_{abba} \\ \chi^{(3)}_{baba} = \chi^{(3)}_{baab} \end{cases},$$

with the first subscript (from left to right) denoting the transition moment for SHG, the second for photon wave vector **q**, and the last two for the fundamental inputs. $\overleftrightarrow{\chi}_{EQ}^{(3)}$ also has the following nonzero tensor elements involving the out-of-plane *c*-axis [1]:

$$\chi_{cccc}^{(3)}, \chi_{aaaa}^{(3)} = \chi_{bbbb}^{(3)}, \chi_{aacc}^{(3)} = \chi_{bbcc}^{(3)}, \chi_{ccaa}^{(3)} = \chi_{ccbb}^{(3)}, \chi_{caac}^{(3)} = \chi_{cbbc}^{(3)}, \chi_{acca}^{(3)} =$$

$$\chi_{bccb}^{(3)}, \chi_{acac}^{(3)} = \chi_{bcbc}^{(3)}, \chi_{caca}^{(3)} = \chi_{cbcb}^{(3)}.$$

The lab (*x*, *y*, *z*) and lattice (*a*, *b*, *c*) coordinates are related by $\hat{a} = \cos\phi\hat{x} + \sin\phi\hat{y}$, $\hat{b} = -\sin\phi\hat{x} + \cos\phi\hat{y}$ and $\hat{c} = \hat{z}$, with $\phi$ being the angle of *x* from the *a-c* plane. Coordinate transformation connects the rank-four nonlinear susceptibility tensors expressed in the lab and lattice coordinates via $\chi_{ijkl}^{(3)} = \sum_{mnop} \chi_{mnop}^{(3)} (\hat{i} \cdot \hat{m})(\hat{j} \cdot \hat{n})(\hat{k} \cdot \hat{o})(\hat{l} \cdot \hat{p})$, where $i, j, k, l \in \{x, y, z\}$ and $m, n, o, p \in \{a, b, c\}$. After the transformation, $\overleftrightarrow{\chi}_{EQ}^{(3)}$ in the lab coordinates is found to be independent of $\phi$ and contains the following nonzero in-plane tensor elements:

$$\chi_{xxxx}^{(3)} = \chi_{yyyy}^{(3)} = \chi_{aaaa}^{(3)}, \begin{cases} \chi_{xxyy}^{(3)} = \chi_{yyxx}^{(3)} = \chi_{aabb}^{(3)} \\ \chi_{xyxy}^{(3)} = \chi_{yxyx}^{(3)} = \chi_{abab}^{(3)} \\ \chi_{xyyx}^{(3)} = \chi_{yxxy}^{(3)} = \chi_{abba}^{(3)} \end{cases}, \begin{cases} \chi_{xyxy}^{(3)} = \chi_{xyyx}^{(3)} = \chi_{abab}^{(3)} \\ \chi_{yxyx}^{(3)} = \chi_{yxxy}^{(3)} = \chi_{baba}^{(3)} \end{cases},$$

The ED-SHG from a surface of the $C_{6v}$ point group is generated through the rank-3 susceptibility tensor, $\overleftrightarrow{\chi}_{ED}^{(2)}$, which has the following nonzero tensor elements:

$$\chi_{ccc}^{(2)}, \chi_{caa}^{(2)} = \chi_{cbb}^{(2)}, \chi_{aac}^{(2)} = \chi_{bbc}^{(2)}, \chi_{aca}^{(2)} = \chi_{bcb}^{(2)}, \text{ and } \chi_{aac}^{(2)} = \chi_{aca}^{(2)}.$$

with the first subscript (from left to right) denoting the transition moment for SHG and the other two for the fundamental inputs. In the lab coordinates, the nonzero $\overleftrightarrow{\chi}_{ED}^{(2)}$ elements are:

$$\chi_{zzz}^{(2)} = \chi_{ccc}^{(2)}, \chi_{zxx}^{(2)} = \chi_{zyy}^{(2)} = \chi_{caa}^{(2)}, \chi_{xxz}^{(2)} = \chi_{yyz}^{(2)} = \chi_{aac}^{(2)}, \chi_{xzx}^{(2)} = \chi_{yzy}^{(2)} = \chi_{aca}^{(2)}, \quad \text{and}$$

$$\chi^{(2)}_{xxz} = \chi^{(2)}_{xzx} = \chi^{(2)}_{aac}.$$

**b) Under different excitation geometries**

Experimentally, the EQ-SHG response is proportional to an effective second-order nonlinear susceptibility, that is:

$$\mathbf{E}_{EQ}(2\omega) \propto \overleftrightarrow{\chi}^{(2)}_{EQ,eff}:\mathbf{E}(\omega)\mathbf{E}(\omega), \text{ and } \overleftrightarrow{\chi}^{(2)}_{EQ,eff} = \overleftrightarrow{L}(2\omega):\left(\overleftrightarrow{\chi}^{(3)}_{EQ} \cdot \mathbf{q}\right):\overleftrightarrow{L}(\omega)\overleftrightarrow{L}(\omega),$$

with $\overleftrightarrow{L}$ being Fresnel coefficients at the corresponding wavelength [1]. At oblique incidence, there is $\mathbf{q} = q_x\hat{x} + q_z\hat{z} = q(\sin\beta\,\hat{x} - \cos\beta\,\hat{z})$, with $\beta$ being the incidence angle. So for the S-polarized input, the output will be P-polarized, and there is:

$$\chi^{(2)}_{EQ,eff}(PSS)\Big|_{obl} =$$
$$-\sin\beta\cos\beta\left(L_{xx}(2\omega)\chi^{(3)}_{EQ,xxyy} + L_{zz}(2\omega)\chi^{(3)}_{EQ,zzyy}\right)L_{yy}(\omega)^2 q(\omega).$$

For the ED-SHG, we have $\mathbf{E}_{ED}(2\omega) \propto \overleftrightarrow{\chi}^{(2)}_{ED,eff}:\mathbf{E}(\omega)\mathbf{E}(\omega)$, and $\overleftrightarrow{\chi}^{(2)}_{ED,eff} = \overleftrightarrow{L}(2\omega):\overleftrightarrow{\chi}^{(2)}_{ED}:\overleftrightarrow{L}(\omega)\overleftrightarrow{L}(\omega)$. At oblique incidence, for the S-in, P-out case, we have:

$$\chi^{(2)}_{ED,eff}(PSS)\Big|_{obl} = \sin\beta\, L_{zz}(2\omega)\chi^{(2)}_{ED,zyy}L_{yy}(\omega)^2.$$

Under the normal incidence, at the focal point of the objective, the incident field must be polarized in-plane, and there is only the $z$ component of the incident photon wave vector, so the EQ-SHG becomes:

$$\chi^{(2)}_{EQ,eff}\Big|_{nor} = -C_{nor}L_{zz}(2\omega)\chi^{(3)}_{EQ,zzxx}\left(L_{xx}(\omega)^2 + L_{yy}(\omega)^2\right)q(\omega).$$

$C_{nor}$ is a geometric factor that would be explained later. For the ED-SHG, there is:

$$\chi^{(2)}_{ED,eff}\Big|_{nor} = C_{nor}L_{zz}(2\omega)\chi^{(2)}_{ED,zxx}\left(L_{xx}(\omega)^2 + L_{yy}(\omega)^2\right).$$

Both generate polarizations at $2\omega$ along the $z$ direction. Using an objective of large enough NA, we can collect its radiation with a good efficiency, defined by the geometric factor that $C_{nor} \approx \frac{1}{2}\int_0^{\sin^{-1}NA} \sin^2\beta'\, d\beta'$. For NA = 0.95, we have $C_{nor} \approx 0.24$. Therefore, with the substrate refractive index taking as 1.45 (neglecting the dispersion) and $\beta = 45°$, we found the ratio between the normal and oblique EQ-SHG is:

$$\gamma_{EQ} = \left|\chi^{(2)}_{EQ,eff}\right|_{nor}/\left.\chi^{(2)}_{EQ,eff}(PSS)\right|_{obl}\Big|^2 =$$

$$\left|0.17\chi^{(3)}_{EQ,zzxx}/(0.24\chi^{(3)}_{EQ,xxyy} + 0.13\chi^{(3)}_{EQ,zzyy})\right|^2 \approx 0,$$

because $\chi^{(3)}_{EQ,xxyy} \gg \chi^{(3)}_{EQ,zzxx} = \chi^{(3)}_{EQ,zzyy}$ for the 2D graphene monolayer. The ratio for ED-SHG is:

$$\gamma_{ED} = \left|\chi^{(2)}_{ED,eff}\right|_{nor}/\left.\chi^{(2)}_{ED,eff}(PSS)\right|_{obl}\Big|^2 \approx 0.96 = 96\%.$$

Experimentally, the intensity ratio between the normal and oblique SHG is only about 0.6%, which is far less than that expected for the ED-SHG. Therefore, the major contribution to the oblique SHG should be of the EQ-type.

### 3. Calculation of the polarization patterns

#### a) EQ-SHG

The electric-quadrupole and/or magnetic-dipole contributions (EQ) [23] is described by the rank-four susceptibility tensor $\overleftrightarrow{\chi}^{(3)}$. The EQ term contributes via $\mathbf{q} \cdot \nabla_{\mathbf{k}}$ in matrix elements, with $\mathbf{q}$, $\mathbf{k}$ being the photon and electron wave vectors, respectively [2]. Therefore, a nonlinear polarization at $2\omega$ can be induced via $\mathbf{P}_{EQ}(2\omega = \omega + \omega) = \overleftrightarrow{\chi}^{(3)}_{EQ}(\omega) \vdots \mathbf{q}\mathbf{E}(\omega)\mathbf{E}(\omega)$, with $\overleftrightarrow{\chi}^{(3)}_{EQ}$ being the EQ-type rank-four susceptibility tensor, and $\mathbf{E}(\omega) = \mathcal{E}e^{i(\mathbf{q}\cdot\mathbf{r}-\omega t)}$ the excitation field at frequency $\omega$. In general, susceptibility tensor elements include matrix elements of both intraband and interband transition dipole moments with energy denominators [1]. Because of the 2D nature of graphene and the dominance of $P_z$ orbital, both the intraband and interband transition matrix elements along the $z$ direction are negligible [23, 31]. Thus for $\overleftrightarrow{\chi}_{EQ}^{(3)}$ tensor, we consider only in-plane subset described in Section 2. At oblique incidence, we have $\mathbf{q}_\parallel = q_x \hat{\mathbf{x}}$ regardless of the input polarization. So for an P-input, the induced polarization is $\mathbf{P}_{EQ}(2\omega) = -\cos^2\beta\, L_{xx}(\omega)^2 \chi_{EQ,xxxx}^{(3)} q_x E(\omega)^2 \hat{\mathbf{x}}$ ($\beta$, $L_{ii}$ being the beam incident angle and Fresnel coefficients [1], respectively) that generates a P-polarized SHG output. For the S-input, the induced polarization is $\mathbf{P}_{EQ}(2\omega) = -L_{yy}(\omega)^2 \chi_{EQ,xxyy}^{(3)} q_x E(\omega)^2 \hat{\mathbf{x}}$, which still corresponds to a P-polarized SHG output. When the input field has both P and S components, the cross terms in $\overleftrightarrow{\chi}_{EQ}^{(3)}$ cause the polarization vector of the output signal to rotate, through:

$$E_{EQ,P}(2\omega) \propto -\cos\beta\, L_{xx}(2\omega) P_{EQ,x}(2\omega) =$$

$$-\cos\beta \sin\beta\, L_{xx}(2\omega)\left(\cos^2\beta \cos^2\theta_0\, L_{xx}(\omega)^2 \chi_{EQ,xxxx}^{(3)} + \sin^2\theta_0\, L_{yy}(\omega)^2 \chi_{EQ,xxyy}^{(3)}\right) qE(\omega)^2,$$

$$E_{EQ,S}(2\omega) \propto L_{yy}(2\omega) P_{EQ,y}(2\omega) =$$

$$2\cos\beta \sin\theta_0 \cos\theta_0 \sin\beta\, L_{yy}(2\omega) L_{yy}(\omega) L_{xx}(\omega) \chi_{EQ,yxyx}^{(3)} E(\omega)^2,$$

Where we used the relation $\chi_{EQ,yxxy}^{(3)} = \chi_{EQ,yxyx}^{(3)}$, and $\theta_0$ denotes the angle between the input polarization vector and the beam incident plane. Since $\overleftrightarrow{\chi}_{EQ}^{(3)}$ terms are in general complex numbers, the phase shift between $E_{EQ,P}(2\omega)$ and $E_{EQ,S}(2\omega)$ can induce elliptically polarized output.

Experimentally, by comparing the signals in Fig. 3(c) and 3(d), we can obtain the ratio $a = \left|\chi^{(3)}_{EQ,xxxx}/\chi^{(3)}_{EQ,xxyy}\right|$. We further define:

$$b = \left|\frac{E_{EQ,S}}{E_{EQ,P}}\right|, \text{ and } \delta = \text{Arg}\left(\frac{E_{EQ,S}}{E_{EQ,P}}\right).$$

When the signal analyzer is set at an angle $\theta$ from the beam incident plane, the detected intensity is related to $\theta$ via:

$$I_{EQ-SHG}(\theta) \propto \left|be^{i\delta}\sin\theta + \cos\theta\right|^2,$$

so both $b$ and $\delta$ can be obtained via fitting the polarization pattern in Fig. 3(e). Meanwhile, all above values can be calculated based on theoretical models by Cheng *et al.* [23] and Wang *et al.* [22]. The comparison between the experimental and theoretical results at $\mu = -0.88$ eV and $\theta_0 = 45°$ are listed in Supplementary Table 1.

|   | Experimental | Theoretical Cheng et al. [23] | Theoretical Wang et al. [22] |
|---|---|---|---|
| $a$ | 0.95 ±0.02 | 0.65 | 0.97 |
| $b$ | 2.09 ±0.06 | 2.39 | 3.08 |
| $\delta$ | 111.2° ±1.5° | 110.3° | 78.2° |

**Supplementary Table 1.** Comparison of the fitting parameters a, b and $\delta$ between the experimental data in Fig. 3(e) and theoretical calculations. Note that the difference between Cheng et al. [23] and Wang et al. [22] arises from the inclusion of the phenomenological relaxation parameter.

In the theoretical calculation, we took the temperature to be at 300 K (for $2|\mu| \gg \hbar\omega$) [14], the resonant damping factor $\Gamma = 0.2\times|\mu|$ eV, the Fresnel coefficients for the air/substrate interface with the substrate refractive index taken as 1.45. It is noted that the µ-dependent peak corresponding to the two-photon Fermi edge resonance ($|\mu| = \hbar\omega$) appeared down-shifted in energy [14]. Thus we used a scaling factor of 1.14 for the chemical potential $\mu$ ($|\mu| = 1.14 \times 0.88$ eV $= 1.00$ eV) in the calculation, which was obtained by matching the experimental and theoretical energies of the two-photon resonance in Fig. 4(a) and 4(c).

**b) ED-THG**

Since THG is electric-dipole (ED) allowed, the leading term is via the following process: $\mathbf{P}_{ED}(3\omega = \omega + \omega + \omega) = \overleftrightarrow{\chi}^{(3)}_{ED}(\omega) \vdots \mathbf{E}(\omega)\mathbf{E}(\omega)\mathbf{E}(\omega)$, with $\mathbf{P}_{ED}$ being the nonlinear polarization at $3\omega$, and $\overleftrightarrow{\chi}^{(3)}_{ED}$ the ED-type rank-four susceptibility tensor [1]. Again due to the 2D nature of graphene, we also consider only in-plane components of $\chi^{(3)}_{ED}$ for THG [14,31]. Therefore, for the S-input, $\mathbf{E}(\omega) \parallel \hat{\mathbf{y}}$, and the only component that could contribute is $\chi^{(3)}_{ED,yyyy}$, so we have $\mathbf{P}_{ED}(3\omega) = L_{yy}(\omega)^3 \chi^{(3)}_{ED,yyyy} E(\omega)^3 \hat{\mathbf{y}}$, and the output field is also S-polarized. For the P-input, $\mathbf{E}(\omega)$ has both $x$ and $z$ components, and the only contributing $\overleftrightarrow{\chi}^{(3)}_{ED}$ component is $\chi^{(3)}_{ED,xxxx}$, which gave rise to $\mathbf{P}_{ED}(3\omega) = \cos^3\beta\, L_{xx}(\omega)^3 \chi^{(3)}_{EQ,xxxx} E(\omega)^3 \hat{\mathbf{x}}$ that gives a P-polarized output. When there are both P and S inputs, we have:

$E_{ED,P}(3\omega) \propto -\cos^2\beta \cos\theta_0\, L_{xx}(3\omega) L_{xx}(\omega) \big(\cos^2\beta \cos^2\theta_0\, L_{xx}(\omega)^2 + \sin^2\theta_0\, L_{yy}(\omega)^2\big) \chi^{(3)}_{ED,xxxx}$,

$E_{ED,S}(3\omega) \propto \sin\theta_0\, L_{yy}(3\omega) L_{yy}(\omega) \big(\cos^2\beta \cos^2\theta_0\, L_{xx}(\omega)^2 + \sin^2\theta_0\, L_{yy}(\omega)^2\big) \chi^{(3)}_{ED,xxxx}$.

In the above derivation we used the permutation relations for the $D_{6h}$ (6/mmm) symmetry group shown in Section 2. The last set of equations shows that $E_{ED,P}(3\omega)$ is of the same phase with $E_{ED,S}(3\omega)$ regardless of the phase of $\overleftrightarrow{\chi}_{ED}^{(3)}$ elements, so the THG output is always linearly polarized. The angle $\theta_{THG}$ between the THG polarization and the incident plane can be calculated via:

$$\theta_{THG} = \tan^{-1}(E_{ED,S}/E_{ED,P}) = \tan^{-1}\left(-\frac{\sin\theta_0}{\cos\theta_0}\frac{L_{yy}(3\omega)L_{yy}(\omega)}{\cos^2\beta L_{xx}(3\omega)L_{xx}(\omega)}\right),$$

which equals 50.2° at $\theta_0 = 45°$. This agrees reasonably with the experimental value of $\theta_{THG} = 47.7° \pm 0.6°$ obtained via fitting the THG polarization pattern in Fig. 3(e).

## 4. Estimation of $|\chi_{EQ}^{(3)}q|$ for EQ-SHG

As discussed above, the intensity of the EQ-SHG with S-polarized input and P-polarized output is:

$$I_{PSS}(2\omega) = \frac{(2\omega)^2}{8\epsilon_0 c^3} L_{xx}(2\omega)^2 L_{yy}(\omega)^4 \left|\chi_{EQ,xxyy}^{(3)} q_x\right|^2 d^2 * I(\omega)^2,$$

where the peak intensity $I(\omega)$ is related to the average power $P(\omega)$ by $I(\omega) = 8\left(\frac{\ln 2}{\pi}\right)^{3/2}\frac{P(\omega)}{f*\tau*W^2}$, assuming pulses with the repetition rate $f$, pulse duration $\tau$ and a Gaussian spatial profile of diameter $W$ on the sample; $\epsilon_0$ is the permittivity of the vacuum and $c$ the speed of the light, and d is the thickness of the graphene. Given the efficiency of the signal collection and detection [14], we obtained the value of $|\chi_{EQ,xxyy}^{(3)} q_x|$ about 30 pm/V, and $|\chi_{EQ,xxyy}^{(3)}|$ about $9\times10^{-18}$ m$^2$/V at $\mu = -0.75$ eV. The effective second-order nonlinear susceptibility $|\chi_{EQ,xxyy}^{(3)} q_x|$ was further confirmed by calibrating with that of hBN, MoS$_2$, WSe$_2$ monolayers and bulk z-cut quartz (Fig. S2) on the same experimental setup. Similarly, the value of $|\chi^{(3)}_{EQ,xxxx} q_x|$ for P-polarized input and P-polarized output was estimated. As shown in Fig. 4(a) and Fig. 4(c), the estimations are close to the theoretical values by Cheng *et al.* [23] and Wang *et al.* [22].

## 5. The symmetry of $\vec{\chi}^{(3)}_{EQ}$ with respect to $\mu$

### a) The time reversal symmetry

Under the time reversal operation $\mathcal{K}$, there is: $\mathcal{K}|s\mathbf{k}\rangle = |s(-\mathbf{k})\rangle$ ($s$ being the band index, and $\mathbf{k}$ the electron wave vector), and $\mathcal{K} v_{\mathbf{k}} \mathcal{K}^{-1} = -v_{-\mathbf{k}}$ for the velocity operator $v_{\mathbf{k}}$. So we have:

$$\mathcal{K} \mathcal{P}_{\mathbf{k}}(\mathbf{q},t) \mathcal{K}^{-1} = -\mathcal{P}_{-\mathbf{k}}(-\mathbf{q},-t)$$

for the EQ-SHG polarization $\mathcal{P}_{\mathbf{k}}(\mathbf{q},t)$ [23], with the incident photon wave vector of $\mathbf{q}$. The detailed expression of $\mathcal{P}_{\mathbf{k}}(\mathbf{q},\omega)$, as the Fourier transformation of $\mathcal{P}_{\mathbf{k}}(\mathbf{q},t)$ with respect to the time t, is given in Ref. 23. Note that we used a different notation from those in Ref. 23, they are related by $\mathcal{P}_{\mathbf{k}}(\mathbf{q},\omega) = \tilde{\mathcal{P}}^{(2)}_{\mathbf{k}}(q,q; 2\hbar\omega + i\gamma, \hbar\omega + i\gamma)$. Define $|s\mathbf{k}\rangle_t = \mathcal{K}|s\mathbf{k}\rangle = |s(-\mathbf{k})\rangle$, we have:

$$[\mathcal{P}_{ss'\mathbf{k}}(\mathbf{q},t)]^* = \langle \mathcal{K} s\mathbf{k}|\mathcal{K}\mathcal{P}_{\mathbf{k}}(\mathbf{q},t)\mathcal{K}^{-1}|\mathcal{K}s'\mathbf{k}\rangle = \langle s(-\mathbf{k})|-\mathcal{P}_{-\mathbf{k}}(-\mathbf{q},-t)|s'(-\mathbf{k})\rangle = -\mathcal{P}_{ss'(-\mathbf{k})}(-\mathbf{q},-t).$$

The Fourier transformation then gives:

$$[\mathcal{P}_{ss'\mathbf{k}}(\mathbf{q},\omega)]^* = -\mathcal{P}_{ss'(-\mathbf{k})}(-\mathbf{q},\omega). \qquad \text{Eq. S1}$$

### b) The electron-hole symmetry

The electron-hole symmetry operator is $\mathcal{C} = \sigma_z$, which gives $\mathcal{C} H_{\mathbf{k}} \mathcal{C}^{-1} = -H_{\mathbf{k}}$, $\mathcal{C}|s\mathbf{k}\rangle = -|\bar{s}\mathbf{k}\rangle$, and $\mathcal{C} v_{\mathbf{k}} \mathcal{C}^{-1} = -v_{\mathbf{k}}$. We then have:

$$\mathcal{C}\mathcal{P}_k(q,t)\mathcal{C}^{-1} = -\mathcal{P}_k(q,-t),$$

which leads to:

$$\mathcal{P}_{ss'k}(q,t) = \langle sk|\mathcal{C}^{-1}\mathcal{C}\mathcal{P}_k(q,t)\mathcal{C}^{-1}\mathcal{C}|s'k\rangle = -\langle sk|\mathcal{C}^{-1}\mathcal{P}_k(q,-t)\mathcal{C}|s'k\rangle$$

$$= -\langle \bar{s}k|\mathcal{P}_k(q,-t)|\bar{s}'k\rangle = -\mathcal{P}_{\bar{s}\bar{s}'k}(q,-t)$$

The Fourier transformation gives:

$$\mathcal{P}_{ss'k}(q,\omega) = -\mathcal{P}_{\bar{s}\bar{s}'k}(q,-\omega). \qquad \text{Eq. S2}$$

c) **The Hermitian**

The Hermitian of $\mathcal{P}_k$ is $[\mathcal{P}_k(q,t)]^\dagger = \mathcal{P}_k(-q,t)$, so there is $[\mathcal{P}_{ss'k}(q,t)]^* = \mathcal{P}_{ss'k}(-q,t)$, and the Fourier transformation gives:

$$[\mathcal{P}_{ss'k}(q,\omega)]^* = \mathcal{P}_{ss'k}(-q,-\omega). \qquad \text{Eq. S3}$$

d) **The symmetry over** $\mu$

Combining Eqs. S1 and S3, we have

$$\mathcal{P}_{ss'k}(q,\omega) = -\mathcal{P}_{ss'(-k)}(q,-\omega),$$

and further with Eq. S2, there is:

$$\mathcal{P}_{ss'k}(q,\omega) = \mathcal{P}_{\bar{s}\bar{s}'(-k)}(q,\omega).$$

The left- and right-hand-side correspond to transitions at opposite sides of the Dirac point, as illustrated in Fig. 4(g) in the main text. Finally, because of the electron-hole symmetry, the band occupation number satisfies:

$$n_{sk}(\mu) = 1 - n_{\bar{s}(-k)}(-\mu),$$

The nonlinear susceptibility is thus:

$$\chi_{EQ}^{(3)}(\boldsymbol{q},\omega,\mu) \propto \sum_{ss'\boldsymbol{k}} \mathcal{P}_{ss'\boldsymbol{k}}(\boldsymbol{q},\omega) n_{s\boldsymbol{k}}(\mu) =$$

$$\sum_{ss'\boldsymbol{k}} \mathcal{P}_{ss'\boldsymbol{k}}(\boldsymbol{q},\omega) - \sum_{ss'\boldsymbol{k}} \mathcal{P}_{\bar{s}\bar{s}'(-\boldsymbol{k})}(\boldsymbol{q},\omega) n_{\bar{s}(-\boldsymbol{k})}(-\mu) =$$

$$\sum_{ss'\boldsymbol{k}} \mathcal{P}_{ss'\boldsymbol{k}}(\boldsymbol{q},\omega) - \chi_{EQ}^{(3)}(\boldsymbol{q},\omega,-\mu).$$

$\sum_{ss'\boldsymbol{k}} \mathcal{P}_{ss'\boldsymbol{k}}(\boldsymbol{q},\omega)$ corresponds to the case with all states occupied and should be zero, so we find:

$$\chi_{EQ}^{(3)}(\boldsymbol{q},\omega,\mu) = -\chi_{EQ}^{(3)}(\boldsymbol{q},\omega,-\mu),$$

which is an odd function with respect to $\mu$. Figure 4(d) and 4(e) show simulated curves of $\chi_{EQ}^{(3)}(\mu)$ at $\hbar\omega = 0.95$ eV and oblique incidence of 45°, and the odd parity over $\mu$ is clearly seen. By comparsion, $\chi_{ED}^{(3)}(\mu)$ is an even function with respect to $\mu$ [Fig. 4(f)].

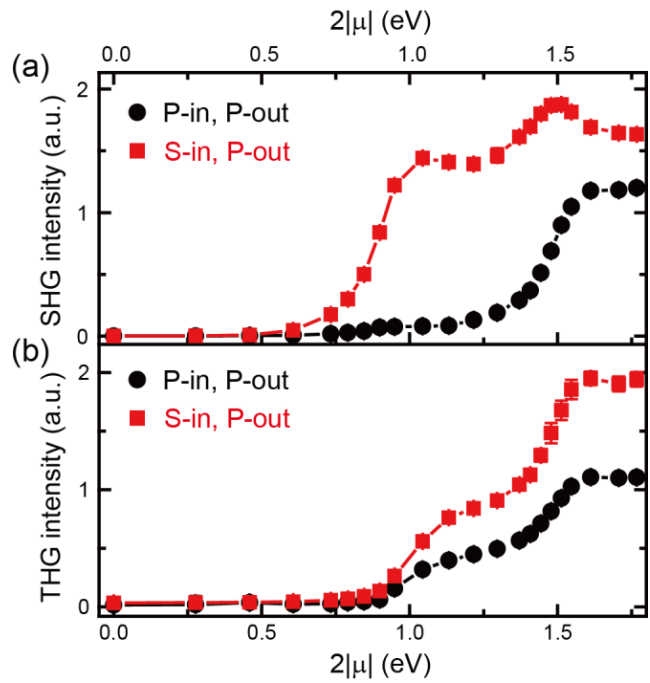

FIG. S1. SHG and THG dependence of chemical potential $\mu$ at oblique incidence of 45 °. The graphene supported on a fused silica was ion-gel gated to tune the chemical potential $\mu$. The data were extracted from Fig. 1(b) and Fig. 1(d) for negative chemical potentials.

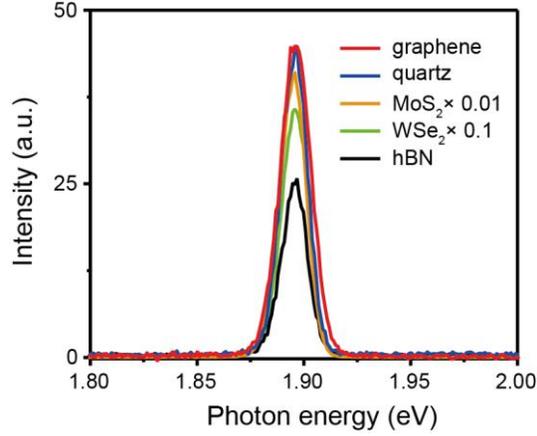

FIG. S2. Comparison of SHG intensity in some common 2D materials. SHG spectra from heavily doped graphene ($\mu \approx -0.75$ eV), hBN, $MoS_2$, $WSe_2$ monolayers, and the bulk z-cut quartz. All the SHG spectra were measured by the same oblique setup (Fig. S3) at the same incident fluence of 2.4 J/m$^2$ (S-polarized excitation). $MoS_2$ and $WSe_2$ monolayers were exfoliated on fused silica substrate. hBN monolayer was grown by chemical vapour deposition (CVD) and transferred onto silicon substrate. The SHG spectra on $MoS_2$ and $WSe_2$ monolayers are scaled down by 100 and 10 times, respectively. Correspondingly, their second order nonlinear susceptibilities are about 10 and 3 times stronger than that of doped graphene.

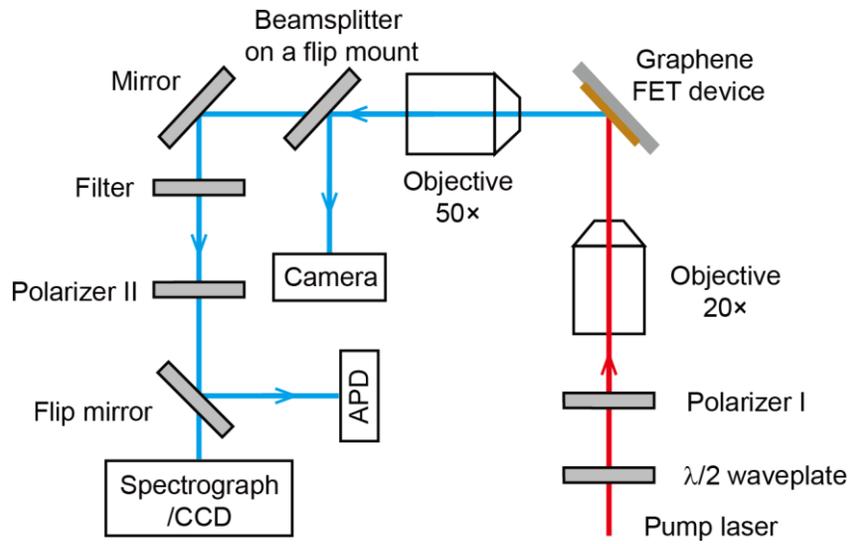

FIG. S3. Schematic of the measurement setup at oblique incidence of 45°. APD: avalanche photodiode. The numerical apertures for the excitation objective (20×) and collection objective (50×) are 0.35 and 0.40, respectively.